\documentstyle[prl,aps] {revtex}                     
\begin{document}
\newcommand{\snp}{S_{0+}^{*}}
\newcommand{\sem}{S_{1-}^{*}}
\newcommand{\sep}{S_{1+}^{*}}
\newcommand{\snpx}{S_{0+}}
\newcommand{\semx}{S_{1-}}
\newcommand{\sepx}{S_{1+}}
\newcommand{\enp}{E_{0+}}
\newcommand{\eem}{E_{1-}}
\newcommand{\eep}{E_{1+}}
\newcommand{\mem}{M_{1-}}
\newcommand{\mep}{M_{1+}}
\newcommand{\EMRpi}{\mbox{EMR}^{\pi^0}}
\newcommand{\EMRiso}{\mbox{EMR}^{3/2}}
\newcommand{\CMRpi}{\mbox{CMR}^{\pi^0}}
\newcommand{\CMRiso}{\mbox{CMR}^{3/2}}
\newcommand{\tEMRpi}{\widetilde{\mbox{EMR}}^{\pi^0}}
\newcommand{\tCMRpi}{\widetilde{\mbox{CMR}}^{\pi^0}}
\newcommand{\tEMRiso}{\widetilde{\mbox{EMR}}^{3/2}}
\newcommand{\tCMRiso}{\widetilde{\mbox{CMR}}^{3/2}}
\newcommand{\tl}{\tilde\lambda}
%
\title{Measurement of the beam-helicity asymmetry in the 
       $p(\vec e, e'p)\pi^0$  reaction at the energy of the $\Delta(1232)$  
       resonance}

\author{
P.~Bartsch$^1$, 
D.~Baumann$^1$,  
J. Bermuth$^2$,
R.~B\"ohm$^1$, 
K.~Bohinc$^{1,3}$, 
D.~Bosnar$^1$\thanks{permanent address: Department of Physics, University of 
                     Zagreb, Croatia}, 
M.~Ding$^1$, 
M. Distler$^1$, 
D.~Drechsel$^1$, 
D. Elsner$^1$, 
I.~Ewald$^1$, 
J.~Friedrich$^1$, 
J.M.~Friedrich$^1$\thanks{present address: Physik Department E18, 
                                           TU M{\"u}nchen, Germany}, 
S.~Gr\"ozinger$^1$, 
S.~Hedicke$^1$,
P.~Jennewein$^1$, 
M.~Kahrau$^1$,
S.S.~Kamalov$^1$\thanks{permanent address: Laboratory for Theoretical Physics,
                                           JINR Dubna, Russia}, 
F.~Klein$^1$, 
K.W.~Krygier$^1$, 
A.~Liesenfeld$^1$, 
H.~Merkel$^1$, 
P.~Merle$^1$,
U.~M\"uller$^1$, 
R.~Neuhausen$^1$, 
Th.~Pospischil$^{1}$,
M.~Potokar$^3$, 
G.~Rosner$^1$\thanks{present address: Dept. of Physics and Astronomy,
                                      University of Glasgow, UK}, 
H.~Schmieden$^1$\thanks{corresponding author, email: hs@kph.uni-mainz.de}, 
M.~Seimetz$^1$, 
A.~S\"ule$^1$, 
L.Tiator$^1$,
A.~Wagner$^1$, 
Th.~Walcher$^1$, 
M. Weis$^1$ 
       }
\address{$^1$  Institut f\"ur Kernphysik, Universit\"at Mainz, D-55099 Mainz,  
               Germany \\                                                    
         $^2$  Institut f\"ur Physik, Universit\"at Mainz, D-55099 Mainz, 
               Germany \\                   
         $^3$  Institut Jo\v zef Stefan, University of Ljubljana,            
               SI-1001 Ljubljana, Slovenia \\                                
        }                                                                    

\date{\today}

\maketitle

\begin{abstract}
In a $p(\vec e, e'p)\pi^0$  out-of-plane coincidence experiment
at the 3-spectrometer setup of the Mainz Microtron MAMI,
the beam-helicity asymmetry has been precisely measured around
the energy of the $\Delta(1232)$  resonance and $Q^2 = 0.2$\,(GeV/c)$^2$.
The results are in disagreement with three up-to-date model calculations.
This is interpreted as lack of understanding of the non-resonant background,
which in dynamical models is related to the pion cloud.
\pacs{PACS numbers: 13.60.Le, 14.20.Gk, 13.40.-f, 13.60.-r}  
\end{abstract}
\twocolumn
%
%

Based on lepton and hadron scattering experiments the nucleon
is considered as composed of quarks and gluons. 
At high energies and large momentum transfers 
this structure can be consistently described in 
terms of perturbative Quantum Chromodynamics, 
because the strong coupling becomes small at the corresponding 
spatial scale, the regime of `asymptotic freedom'.
In contrast, at distances of the size of the nucleon perturbative methods
fail.
Therefore, it is still an open question how QCD generates the observed 
`confinement' of the quarks.
At the nucleon-size scale, low momentum-transfer experiments can help 
our understanding of the confinement mechanism by testing QCD-motivated 
models.

A direct consequence of the nucleon's substructure 
in the confinement regime is its excitation spectrum.
To study the underlying internal dynamics,
the prominent first excited state, 
the $\Delta(1232)$  resonance with spin and isospin 3/2, 
has been extensively studied with both hadronic and electromagnetic probes.
In the naive quark model it emerges from the ground state by the
spin flip of one of the constituent quarks, a pure M1 transition where
one unit of angular momentum $\Delta L = 1$  is transferred.
In contrast to pion scattering, 
electromagnetic excitation in principle accesses the positive
parity $\Delta(1232)$  with both $\Delta L = 1$  and $2$.
While not existing in the naive quark model,
quadrupole $\Delta L = 2$  transitions become possible through d-state
admixtures in the baryon wave function,
which in QCD-motivated constituent quark models are generated by the color 
hyperfine interaction between the quarks 
\cite{deRujula75,IKK82,GD82,DG84}.

However, the measured electric and scalar quadrupole to magnetic dipole
ratios of $R_{EM} \simeq -2.5\,\%$  and $R_{SM} \simeq -6.5\,\%$  
\cite{Beck97,Blanpied97,Frolov99,Pospischil01,Mertz01,Gothe00,Joo01},
respectively, are up to an order of magnitude larger than predicted 
by those models.
More quadrupole strength is expected in models which emphasize the role
of pions \cite{Gellas99,Silva00,Buchmann98,KY99,SL01}.
Through pion rescattering at the real or virtual photon
$\gamma^{(*)} N \Delta$  vertex, 
the pion cloud is explicitly treated in the dynamical models \cite{KY99,SL01}.
They yield a consistent decomposition into the ``bare'' $\Delta$, 
as described in quark models, and the ``dressing'' by the pion cloud, 
both for the quadrupole ratios and the M1 strength.

In one-photon exchange approximation the fivefold differential cross section
of pion electroproduction,
\begin{equation}
\frac{d^5\sigma}{dE_e d\Omega_e d\Omega_{\pi}^{\rm cm}} = 
                 \Gamma \frac{d^2\sigma_v}{d\Omega_{\pi}^{\rm cm}},
\end{equation} 
factorizes into the virtual photon flux, 
\begin{equation}
\Gamma = \frac{\alpha}{2 \pi^2} \frac{E'}{E} 
         \frac{k_{\gamma}}{Q^2} \frac{1}{1-\epsilon},
\end{equation} 
and the virtual photon cm cross section, 
$d^2\sigma_v/d\Omega_{\pi}^{\rm cm}$.
$\alpha$  denotes the fine structure constant,
$k_\gamma = (W^2 - m_p^2)/2m_p$  the real photon equivalent laboratory energy 
for the excitation of the target with mass $m_p$  to the cm energy $W$, and
$\epsilon = [ 1 + ( 2 |\vec q|^2/Q^2 ) \tan^2{\frac{\vartheta_e}{2}} ]^{-1}$
the photon polarization parameter.
$Q^2 = |\vec q|^2 - \omega^2$  is the squared four-momentum transfer,
$\vec q$  and $\omega$  are the three-momentum and energy transfer, 
respectively, and $E$, $E'$  and $\vartheta_e$  
the incoming and outgoing electron energy, 
and the electron scattering angle in the laboratory frame.

Without target or recoil polarization, the virtual photon cross section 
is given by \cite{DT92}
\begin{eqnarray}
\frac{d^2\sigma_v}{d\Omega_\pi^{cm}} &=& \lambda \cdot
        [      R_T + \epsilon_L R_L 
               + \sqrt{2 \epsilon_L (1+\epsilon)} R_{LT} \cos \Phi +
               \nonumber \\ & &
               \epsilon R_{TT} \cos 2\Phi
               + P_e \sqrt{2 \epsilon_L (1-\epsilon)} R_{LT'} \sin \Phi
        ].
\label{eq:x-sec}
\end{eqnarray}
The factor $\lambda = |\vec p_\pi^{\,\rm cm}|/k_\gamma^{\rm cm}$  
is determined by the pion cm momentum $\vec p_{\pi}^{\,\rm cm}$  and 
$k_{\gamma}^{\rm cm} = (m_p/W) k_{\gamma}$.
The structure functions $R_i$  describe the response of the hadronic 
system to the various polarization states of the photon field, 
which are described by the transverse and longitudinal
polarization, 
$\epsilon$  and $\epsilon_L = \frac{Q^2}{\omega^2_{\rm cm}} \epsilon$, 
respectively, and by the longitudinal electron polarization, $P_e$.
The tilting angle between the electron scattering plane and the reaction plane
is denoted by $\Phi$, where $\Phi = 0$  and $180\,\deg$  correspond to
pions ejected in the electron scattering plane, 
and $\Phi = 90$  and $270\,\deg$ 
perpendicularly to the scattering plane.

High sensitivity to $R_{EM}$  and $R_{SM}$  is obtained
through the pion p-wave interferences 
$\Re{e}\{E_{1+}^*M_{1+}\}$  and $\Re{e}\{S_{1+}^*M_{1+}\}$  occuring in
$R_{TT}$  and $R_{LT}$, respectively.
These interferences
can also be accessed by measuring the recoil
polarization in the $p(\vec e, e'\vec p)\pi^0$  reaction \cite{HS98}.
For parallel kinematics the beam-helicity 
independent polarization component, $P_y$,
reads in s and p wave approximation,
which is only used for simplicity,
\begin{equation}
\sigma_0 P_y =      - c_+ \,\tl ~ 
                    \Im{m}\{(4 S_{1+} + S_{1-} - S_{0+})^* M_{1+}\},
\label{eq:P_y_sp}
\end{equation}
when only terms involving $M_{1+}$  are retained.
$\sigma_0$ denotes the unpolarized cross section,
$c_\pm = \sqrt{2\epsilon_L(1\pm\epsilon)}$  
and $\tl = \omega_{cm}/|\vec q_{cm}|$. 
The experimental results for $P_y$  are not well reproduced by model
calculations \cite{Pospischil01,Warren98}.

It is unclear whether this disagreement originates from higher 
resonances, like the Roper N(1440) which couples to the $S_{1-}$  partial
wave, or to non-resonant contributions.
Moreover, a model independent relation between the transverse and longitudinal
recoil polarization components \cite{ST00} is possibly violated 
by the MAMI experiment \cite{Pospischil01}.

In the s and p wave $\mep$-dominance approximation
the structure function $R_{LT'}$  has a similar
structure as $\sigma_0 P_y$:
\begin{equation}
R_{LT'} = -\sin{\Theta}\,\tl\,\Im{m}\{\,(6\cos{\Theta}\sepx+\snpx)^*\mep \,\}.
\label{eq:R_LTprime}
\end{equation}
The measurement of 
\begin{equation}
\rho_{LT'} = \frac{c_- \,R_{LT'} \,\sin{\Phi}}
                  {R_T + \epsilon_L R_L 
                   + c_+ R_{LT} \cos \Phi 
                   + \epsilon R_{TT} \cos 2\Phi} 
\end{equation}
is thus expected to shed more light on the above discrepancies.
This quantity is experimentally easy to access as an asymmetry with regard to 
the helicity reversal of the electron beam, once out-of-plane proton 
detection is provided \cite{Papanicolas01}.


The $p(\vec e, e'p)\pi^0$  experiment was carried out at the
3--spectrometer facility \cite{Blomqvist98} of the A1 collaboration at 
the Mainz Microtron MAMI.
A typically 6\,$\mu$A electron beam with 80\,\% polarization impinged on 
a 5\,cm long liquid hydrogen target cell made of a
10\,$\mu$m  Havar foil.
Longitudinal beam polarization at the target was obtained by
fine-tuning the beam energy to $E = 854.5$\,MeV.
The beam polarization was measured on a daily basis with a M{\o}ller
polarimeter \cite{Bartsch01} located 15\,m straight upstream of the target.
The scattered electrons were detected in Spectrometer A of the 3--spectrometer
setup, which was set to an angle of $44.5\,\deg$  and a central momentum of
408\,MeV/c.
\input{psfig}
\begin{figure}
\centerline{\psfig{figure=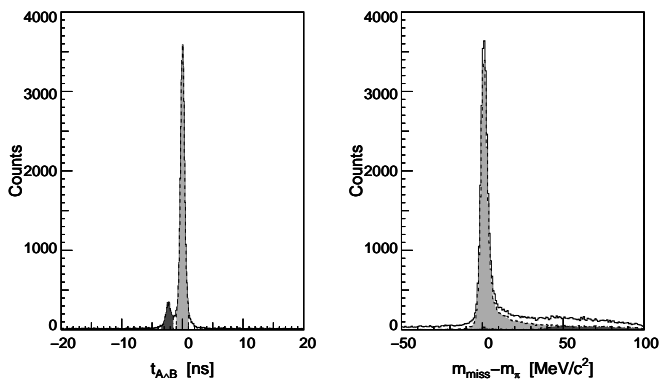,width=8.5cm}} 
\vspace{0.3cm}
\caption{ 
          Left: 
          Coincidence time spectrum between Spectrometer A and B.
          Right:
          Missing mass for the $p(\vec e,e'p)X$  reaction.
          The (hardly visible) dark shaded areas are due to misidentified 
          double pion production, 
          the light shaded areas represent the true $e'p$  coincidences.
          See text for discussion.
        }
\label{fig:time}
\end{figure}            
For the coincident proton detection the out-of-plane capability of
Spectrometer B was used.
It was set to $-26.9\,\deg$  in the horizontal plane and then tilted out of
plane in three different settings of
$\Theta_{\mathrm OOP} = 2$, $7$, and $10\,\deg$.

Both spectrometers are equipped with two double planes of vertical drift
chambers for particle tracking, and two segmented planes of fast plastic
scintillators for particle identification via $dE/dx$  and timing measurements.
The standard detector packages are completed by a threshold gas Cherenkov
detector for $e^\pm$  identification.
In Spectrometer A this device was replaced by the focal plane proton
polarimeter \cite{Pospischil01a} for other experiments.

Figure\,\ref{fig:time}\,(left) shows the coincidence time 
between Spectrometer A (start) and B (stop).
Two prompt peaks are obtained on a tiny random background.
The left peak is associated with $p (e, \pi^- p) \pi^+ e'$
double pion production, where the $\pi^-$  is detected in Spectrometer A 
instead of the scattered electron.
The right peak is well separated.
It represents the true $e'p$  coincidences of the 
$p(\vec e, e'p)\pi^0$  reaction.
A coincidence time resolution of $0.8$\,ns FWHM results in a true to
random ratio of $76:1$.

The final state $\pi^0$  remains unobserved. 
Due to the complete kinematics, the $p(\vec e, e'p)\pi^0$  reaction 
can be reconstructed via the missing mass.
Figure\,\ref{fig:time} (right) shows a clear peak of 
$\simeq 4.5$\,MeV/c$^2$  FWHM at the $\pi^0$  mass.
The strength at higher missing mass is due to random background,
which still is included,
radiative processes and misidentified double pion production.
The latter contribution is marked by the dark shaded areas.
The light shaded areas are related to true coincidences.
A cut of $-5...100$\,MeV/c$^2$  around the $\pi^0$  mass selects the 
$p(e,e'p)\pi^0$  reaction.
There is less than $0.1\,\%$  background remaining after random coincidences
are subtracted.


The beam-helicity asymmetry 
$\rho_{LT'}^{\mathrm exp} = \frac{1}{P_e}(N^+ - N^-) / (N^+ + N^-)$  
is constructed from the numbers of events, $N^\pm$, selected for beam
helicity $+$  and $-$, respectively.
Results for individual bins over the total accepted phase space,
$W                       = 1180...1290 $\, MeV,
$Q^2                     = 0.14...0.26 $\, (GeV/c)$^2$,
$\epsilon                = 0.536 ...0.664 $  and
$\Theta_\pi^{\mathrm cm} = 130...180\,\deg $,   
were obtained, with a varying azimuthal acceptance of
$\Delta\Phi              = 20...360\,\deg $  depending on 
$\Theta_\pi^{\mathrm cm}$.
The stability of the results was checked by varying the cuts in coincidence
time and missing mass.
The latter produced a systematic variation of $\rho_{LT'}$  which, however,
could be entirely attributed to the corresponding variation in the
non-independent other kinematic variables.
The remaining effect of radiative processes is thus estimated to $<1\,\%$.

The systematic errors of the asymmetry measurement are compiled in 
Table\,\ref{tab:syst_errors}.
They are dominated by the uncertainty of the beam polarization.
Individual beam polarization measurements achieve $~ 2\,\%$  accuracy when
statistical and systematical errors are added in quadrature.
Undetected helicity fluctuations are accounted for by an additional 2\,\%
error, which is estimated from long term stability measurements during
$G_E^n$  experiments.

\begin{figure}
\centerline{\psfig{figure=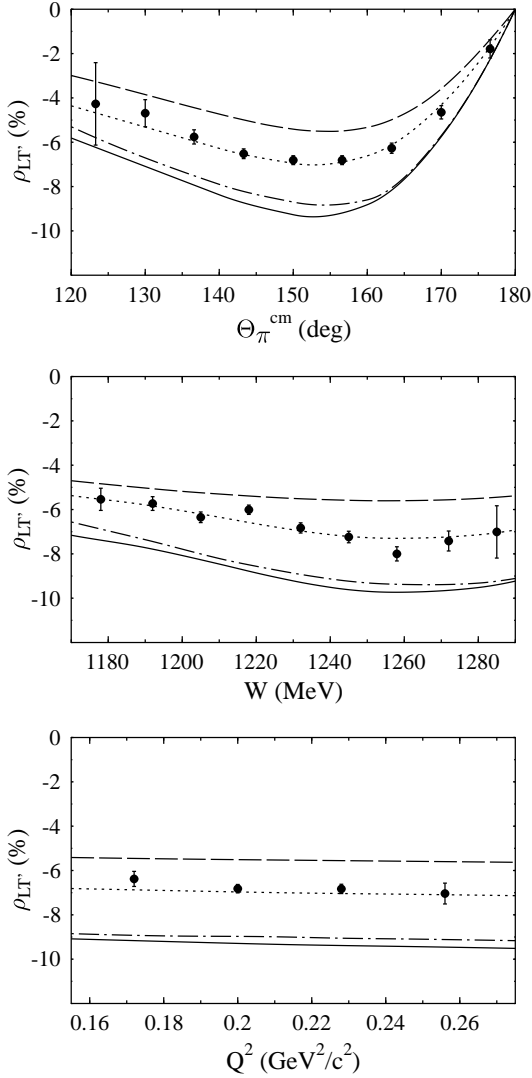,width=8.5cm}}  
\caption{ Results for $\rho_{LT'}$  as a function of 
          $\Theta_\pi^{\mathrm cm}$  (top),
          $W$  (middle) and $Q^2$  (bottom).
          The full curve represents the MAID calculation
          {\protect \cite{Drechsel99}}, the dotted curve is MAID scaled by 
          a factor $0.75$. The dashed and dashed-dotted curves are
          the results of the dynamical models of Sato-Lee
          {\protect \cite{SL01}} and Kamalov-Yang 
          {\protect \cite{KY99}}, respectively.
          Errors are purely statistical.
        }
\label{fig:results}
\end{figure}

For the compilation of the results in Table\,\ref{tab:results} and the 
presentation in Figure\,\ref{fig:results},
$\rho_{LT'}(W,Q^2,\epsilon,\Theta_\pi^{\mathrm cm},\Phi)$  
is projected to nominal kinematics 
($W = 1232$\,MeV, $Q^2 = 0.2$\,(GeV/c)$^2$, $\epsilon = 0.6$, 
 $\Theta_\pi^{\mathrm cm} = 155\,\deg$, $\Phi = 270\,\deg)$  
using the unitary isobar model MAID2000 \cite{Drechsel99}:
\begin{table}
  \begin{center}
    \begin{tabular}{|l|c|}
      error source          & relative error in \% \\
      \hline \hline
      helicity-specific     &             \\
      ~~~luminosity fluctuations
                            & $< 0.5$     \\    
      detector inefficiencies & --        \\    
      background reactions  & $< 0.1$     \\    
      radiative corrections & $< 1.0$     \\    
      beam polarization     & 2.6         \\    
      model uncertainty     & $1.8$       \\    
      \hline
      total (added in quadrature)  & $< 3.4$     \\
      \hline 
    \end{tabular}
  \end{center}
  \caption{Contributions to the systematic error of the measured beam-helicity
      asymmetry. 
      }
  \label{tab:syst_errors}
\end{table}
\begin{equation}
\rho_{LT'} = \frac{\rho_{LT'}^{\mathrm MAID}({\mathrm nom.kin.})}
                      {\rho_{LT'}^{\mathrm MAID}(W,Q^2,\epsilon,
                                       \Theta_\pi^{\mathrm cm},\Phi)}
                 \cdot \rho_{LT'}^{\mathrm exp}(W,Q^2,\epsilon,
                                       \Theta_\pi^{\mathrm cm},\Phi).
\label{eq:projection}
\end{equation}
This is done simultaneously for all except the respective running variables.
An additional systematic error of $1.8\,\%$  due to the projection procedure
is estimated by a $\pm 5\,\%$  variation of the $M_{1+}$  multipole 
in MAID and a $\pm 50\,\%$  variation of the other s and p wave multipoles
(cf. Table\,\ref{tab:syst_errors}).

The results are compared to MAID2000 and the dynamical models of
Kamalov and Yang \cite{KY99} and Sato and Lee \cite{SL01}.
Very similarly to the normal component $P_y$  of the recoil 
proton polarization 
in the $p(\vec e, e'\vec p)\pi^0$  reaction \cite{Pospischil01,Warren98} 
MAID overestimates the magnitude of the asymmetry by one third.  
The appropriately scaled MAID curve describes the differential
dependencies of $\rho_{LT'}$  very well.
This is important for the projection to nominal kinematics 
(Eq.\,\ref{eq:projection}).
While the dynamical model of Kamalov-Yang overestimates $\rho_{LT'}$  
in magnitude as well,
the Sato-Lee model underestimates this quantity.

At present, it therefore appears that neither of the models is capable 
of reproducing the measured $\rho_{LT'}$.
This may be related to the strength of $\snpx$.
MAID2000 simultaneously describes $P_y$  \cite{Pospischil01} 
and the angular distribution of $\rho_{LT'}$ (Fig.\,\ref{fig:results} top),
if the $\Re{e}\,\snpx$  strength is artificially reduced by approximately
60\,\%.
Such contributions can be obtained from pion loops and/or 
dispersion integrals from higher s-wave resonances.
The non-resonant contributions of higher partial waves are more reliably
described by Born terms.
Around $W=1232$\,MeV,
significant non-Born contributions are almost excluded by experiment 
\cite{Joo01,Beck00}.
The Kamalov-Yang model reproduces $P_y$  pretty well, 
but fails at the same time for $\rho_{LT'}$.
It seems that pion cloud effects are not yet consistently included
in the dynamical models.
\begin{table}
  \begin{center}
    \begin{tabular}{|c|c|c|} 
      running variable          & $\rho_{LT'}$ & stat. error \\
      \hline 
      $\Theta_\pi^{\mathrm cm}$  ($\deg$) &    &             \\
      \hline
         123.3  &  -0.0427  &  0.0186  \\
         130.0  &  -0.0469  &  0.0061  \\
         136.6  &  -0.0576  &  0.0032  \\
         143.3  &  -0.0652  &  0.0022  \\
         150.0  &  -0.0681  &  0.0020  \\
         156.6  &  -0.0681  &  0.0020  \\
         163.3  &  -0.0627  &  0.0023  \\
         170.0  &  -0.0465  &  0.0030  \\
         176.6  &  -0.0179  &  0.0041  \\
      \hline
      $W$ (MeV) &           &          \\
      \hline
         1178.  &  -0.0554  &  0.0050  \\
         1192.  &  -0.0573  &  0.0031  \\
         1205.  &  -0.0635  &  0.0024  \\
         1218.  &  -0.0601  &  0.0021  \\
         1232.  &  -0.0683  &  0.0023  \\
         1245.  &  -0.0724  &  0.0026  \\
         1258.  &  -0.0800  &  0.0032  \\
         1272.  &  -0.0742  &  0.0045  \\
         1285.  &  -0.0701  &  0.0118  \\
      \hline
  $Q^2$  (GeV$^2$/c$^2$) &  &          \\
      \hline
         0.172  &  -0.0638  &  0.0034  \\
         0.200  &  -0.0682  &  0.0019  \\
         0.228  &  -0.0683  &  0.0021  \\
         0.256  &  -0.0704  &  0.0047  \\
    \end{tabular}
  \end{center}
  \caption{Results for the beam-helicity asymmetry $\rho_{LT'}$  with
           statistical errors. Except for the respective running variable
           a projection to nominal kinematics has been performed using
           MAID2000 (see text).
      }
  \label{tab:results}
\end{table}
A consistent understanding of the non-resonant background
is an important issue already in the case of the $\Delta(1232)$  
resonance, where this might --- but does not necessarily --- affect the
extraction of resonance properties from $\pi^0$  electroproduction
experiments \cite{HS01}.
It will be mandatory 
for the investigation of higher, weak and overlapping,
resonances.

In summary, for the first time a measurement of the $\rho_{LT'}$  
helicity asymmetry in a $p(\vec e,e'p)\pi^0$  out-of-plane coincidence
experiment is reported. 
The high statistical accuracy is complemented by a small relative systematic 
error which is estimated to be $< 3.4\,\%$.
Neither of three up-to-date model calculations is capable of quantitatively
reproducing the observed asymmetries.
From the failure of the dynamical models it is concluded that pion cloud
effects are not yet sufficiently well understood.

We are indebted to K.-H. Kaiser and K. Aulenbacher and
their staff for providing the excellent polarized beam.
We thank T. Sato and T.-S. H. Lee for providing us with their calculations.
Helpful discussions with R. Beck are gratefully acknowledged.
This work was supported in part by the 
{\em Deutsche Forschungsgemeinschaft} (SFB443) and the federal state
of {\em Rheinland-Pfalz}.


\begin{thebibliography}{99}
\bibitem{deRujula75}    A. de R\'{u}jula, H. Georgi, S.L. Glashow, 
                        Phys. Rev. {\bf D 12}, 147 (1975)
\bibitem{IKK82}         N. Isgur, G. Karl, and R. Koniuk,
                        Phys. Rev. {\bf D 25}, 2394 (1982)
\bibitem{GD82}          S.S. Gershtein and G.V. Dzhikiya,
                        Sov. J. Nucl. Phys. {\bf 34}, 870 (1982)
\bibitem{DG84}          D. Drechsel and M.M. Giannini,
                        Phys. Lett. {\bf 143B}, 329 (1984)
\bibitem{Beck97}        R. Beck et al., Phys. Rev. Lett. {\bf 78}, 606 (1997)
\bibitem{Blanpied97}    G. Blanpied et al., 
                        Phys. Rev. Lett. {\bf 79}, 4337 (1997)  
\bibitem{Frolov99}      V.V. Frolov et al., 
                        Phys. Rev. Lett. {\bf 82}, 45 (1999)
\bibitem{Pospischil01}  Th. Pospischil et al.,
                        Phys. Rev. Lett. {\bf 86}, 2959 (2001) 
\bibitem{Mertz01}       C. Mertz et al.,
                        Phys. Rev. Lett. {\bf 86}, 2963 (2001) 
\bibitem{Gothe00}       R.W. Gothe, 
                        Prog. Part. Nucl. Phys. {\bf 44}, 185 (2000)
\bibitem{Joo01}         K. Joo et al., nucl-ex/0110007v2
\bibitem{Gellas99}      G.C. Gellas et al.,
                        Phys. Rev. \textbf{D 60}, 054022 (1999)
\bibitem{Silva00}       A. Silva et al.,
                        Nucl. Phys. \textbf{A 675}, 637 (2000)
\bibitem{Buchmann98}    A.J. Buchmann et al.,
                        Phys. Rev. {\bf C 58}, 2478 (1998)
\bibitem{KY99}          S.S. Kamalov and S.N. Yang,
                        Phys. Rev. Lett. \textbf{83}, 4494 (1999)
\bibitem{SL01}          T. Sato and T.-S.H. Lee,
                        Phys. Rev. {\bf C 63}, 055201 (2001)
\bibitem{DT92}          D. Drechsel and L. Tiator,
                        J. Phys. G {\bf 18}, 449 (1992)
\bibitem{HS98}          H. Schmieden, Eur. Phys. J. \textbf{A 1}, 427 (1998)
\bibitem{Warren98}      G. Warren et al., Phys. Rev. {\bf C 58}, 3722 (1998)
\bibitem{ST00}          H. Schmieden and L. Tiator, 
                        Eur. Phys. J. \textbf{A 8}, 15 (2000) 
\bibitem{Papanicolas01} C. Papanicolas,
                        proceedings of NSTAR2001, ed. by 
                        D. Drechsel and L. Tiator,
                        World Scientific (2001), p. 11
\bibitem{Blomqvist98}   K.I. Blomqvist et al., 
                        Nucl. Instr. Methods {\bf A 403}, 263 (1998)
\bibitem{Bartsch01}     P. Bartsch, doctoral thesis, Mainz (2002)
                        and P. Bartsch et al., to be published
\bibitem{Pospischil01a} Th. Pospischil et al., 
                        accepted for publication in Nucl. Intr. Meth. {\bf A}
\bibitem{Drechsel99}    D. Drechsel, O. Hanstein, S.S. Kamalov and L. Tiator,
                        Nucl. Phys. {\bf A 645}, 145 (1999)  and
                        http://www.kph.uni-mainz.de/MAID/maid2000/
\bibitem{HS01}          H. Schmieden,
                        proceedings of NSTAR2001, ed. by 
                        D. Drechsel and L. Tiator,
                        World Scientific (2001), p. 27
\bibitem{Beck00}        R. Beck et al.,
                        Phys. Rev. {\bf C 61}, 035204 (2000)
\end{thebibliography}
\end{document}